\documentclass[epj,nopacs]{svjour}
\usepackage{graphicx}
\usepackage{amsmath}

\newcommand{\be}{\begin{equation}}
\newcommand{\ee}{\end{equation}}
\newcommand{\bea}{\begin{eqnarray}}
\newcommand{\eea}{\end{eqnarray}}
\newcommand{\bvec}[1]{\mbox{\boldmath $#1$}}

\begin{document}
\title{Systematics of flux tubes
in the dual Ginzburg-Landau theory and
Casimir scaling hypothesis: folklore and lattice facts}
\author{Yoshiaki Koma\inst{1} \and Miho Koma (Takayama)\inst{2}% etc
\thanks{The family name has changed in May 2001 due to marriage.
}% 
}                     % Do not remove
%
% \offprints{}          % Insert a name or remove this line
%
\institute{Institute for Theoretical Physics, Kanazawa University,
Kanazawa, Ishikawa 920-1192, Japan
\and
Research Center for Nuclear Physics (RCNP), 
Osaka University, 
Mihogaoka 10-1, Ibaraki, Osaka 567-0047, 
Japan}
\date{Received: date / Revised version: date}
\abstract{
The ratios between string tensions $\sigma_D$ of color-electric 
flux tubes in higher and fundamental SU(3) representations, 
$d_{D} \equiv \sigma_{D}/\sigma_{\bf 3}$, 
are systematically studied in a Weyl symmetric formulation of the DGL theory.  
The ratio is found to depend on the Ginzburg-Landau (GL) parameter, 
$\kappa \equiv m_{\chi}/m_{B}$, the mass ratio between monopoles 
($m_{\chi}$) and dual gauge bosons ($m_{B}$).  
While the ratios $d_{D}$ follow a simple flux counting rule in the 
Bogomol'nyi limit, $\kappa=1.0$, systematic
deviations appear with increasing $\kappa$ due to interactions
between fundamental flux inside a higher representation flux tube.  
We find that in a type-II dual superconducting vacuum near $\kappa =
3.0$ this leads to a consistent description of the ratios $d_{D}$ 
observed in lattice QCD simulations.
\PACS{
      {12.38.Aw}{describing text of that key}   \and
      {12.38.Lg}{describing text of that key}
     } % end of PACS codes
} %end of abstract
\authorrunning{Y. Koma and M. Koma}
%% \titlerunning{Systematics of the flux tube
%% in the dual Ginzburg-Landau theory and the Casimir  \ldots}
\titlerunning{Systematics of flux tubes
in the dual Ginzburg-Landau theory and
Casimir \ldots}
\maketitle
\section{Introduction}
\label{intro}
\sloppypar{
The study of the static potential in QCD between
color charges in various representations of the color group SU(3) is
expected to discriminate between possible candidate confinement 
scenarios, {\it i.e.} between the sort of non-perturbative 
vacuum proposed~\cite{Bernard:1983my,Ambjorn:1984mb}.
Recently, such static potentials have been investigated within SU(3) pure
lattice gauge theory extracted from Wilson loops in various
representations $D$~\cite{Deldar:1999vi,Bali:2000un}.  
In Ref.~\cite{Deldar:1999vi} the static potentials have been
parametrized as a superposition of Coulomb, linear and constant terms, 
where the ratios among the string tensions governing the linear terms
were found roughly equal to the ratios between the eigenvalues of the 
quadratic Casimir operator $C^{(2)}(D) = \langle D
| \sum_{a=1}^{8}T^a T^a | D \rangle$ in the respective representation.
This has been considered as a confirmation of the Casimir scaling
hypothesis which is under discussion since 
long~\cite{Bernard:1983my,Ambjorn:1984mb,Faber:1998rp}.  
In Ref.~\cite{Bali:2000un}, instead, the ratios of the static potentials 
themselves is analyzed, which show very good agreement with the 
Casimir ratios uniformly up to distances of $r \sim 1$ fm. 
Following these results, strong arguments have been raised stressing
that QCD-vacuum models should reproduce Casimir 
scaling~\cite{Shevchenko:2000du,Shevchenko:2001ij}.
}

\par 
Recently, in Ref.~\cite{Koma:2001ut}, one of the present authors
has examined the dual Ginzburg-Landau (DGL) 
theory~\cite{Suzuki:1988yq,Maedan:1988yi,Suganuma:1995ps} 
as a QCD-vacuum model under this aspect.
In the DGL theory, the quark confinement mechanism is described by the formation 
of a color-electric flux-tube due to a dual Meissner effect.
The ratios between string tensions of charges in higher and fundamental SU(3)
representations, $d_{D} \equiv \sigma_{D}/\sigma_{\bf 3}$, 
have been calculated within the DGL theory cast into a Weyl symmetric
formulation~\cite{Koma:2000wn,Koma:2000hw}.  
It has been shown that the ratios depend on the GL parameter, 
$\kappa \equiv m_{\chi}/m_{B}$. 
If one wants to reproduce the {\it exact} Casimir ratios $d_D$,
one has to choose the GL parameters depending on representations:
$\kappa \sim 5.0$ for $D={\bf 8}$, ${\bf 6}$,
$\kappa \sim 7.0$ for $D={\bf 10}$, ${\bf 15a}$, 
and 
$\kappa \sim 9.0$ for $D={\bf 27}$, ${\bf 24}$, 
and ${\bf 15s}$.
There was no {\it unique} GL parameter which would have reproduced
{\it all} Casimir ratios at once.

\par
Does this result rule out the DGL theory as a QCD-vacuum model ?
Before hastily drawing this conclusions, we would like to reconsider 
the lattice data.
We find that the string tensions shown in 
Ref.~\cite{Deldar:1999vi} do {\it not obey exact} Casimir scaling. 
They have huge errors for higher representations.
In  Ref.~\cite{Bali:2000un} a clear signal of 
Casimir scaling for \emph{static potentials} are obtained. 
However, it is not clear that this result carries over to the string tensions.
This depends on how one separates the string tension
from the potential, since the tail of the short range Coulomb potential
contributes to the apparent slope of the long range part of the 
potential.
Hence, there is no reason that the ratio between string tensions
is identical with that between potentials.
On the theoretical side, it is not obvious that such group theoretical 
scaling appears for the string tension at distances where  
nonperturbative effects start to become important.

\par
If the behavior of the ratio would be governed exclusively by the group 
theoretical factor, it would be natural to expect that Casimir scaling 
is manifest in {\it arbitrary} SU($N$) gluodynamics. 
However, very recent studies of $k$-strings in SU(4) and SU(6) lattice 
gauge theories did not provide support for Casimir scaling but instead
favored a sine-formula scaling~\cite{DelDebbio:2001kz,Lucini:2001nv}.

\par
In the present paper we study the \emph{systematics} of the ratios
of string tensions among various representations, restricted to
SU(3) gluodynamics, in the DGL theory following Ref.~\cite{Koma:2001ut}.
We want to see whether a certain unique value of the GL parameter can 
provide all the ratios of string tensions consistent with lattice data 
themselves, without any bias toward Casimir scaling. 
Using the method of Ref.~\cite{Koma:2001ut} we can compare the DGL theory 
with the lattice data of Ref.~\cite{Deldar:1999vi}, because 
only there string tensions have been extracted.
Finally, we would like to speculate which features of nonperturbative 
dynamics could be the origin of the representation dependence of the 
string tensions observed in lattice simulations.

\section{The DGL theory}
\sloppypar{
We briefly review how to calculate the string tensions in 
the DGL theory~\cite{Koma:2001ut}.
In order to treat the charges in various SU(3)
representations systematically, we start from the Weyl symmetric form
of the DGL theory~\cite{Koma:2000wn,Koma:2000hw}
\begin{eqnarray}
    \mathcal{L}_{\rm
    DGL} &=& \sum_{i=1}^3 \Biggl [ - \frac{1}{6g^{2}}
    \left ( (\partial    \wedge B_{i})_{\mu\nu} +2\pi \sum_{j=1}^{3} m_{ij}
    \Sigma_{j\;\mu\nu}^{(e)} \right )^{2}
    \nonumber\\*
    && + \left | \left
    (\partial_{\mu} + i B_{i\;\mu}\right )\chi_{i} \right |^2 
    - \lambda \left ( \left |\chi_{i} \right |^2-v^2 \right )^2 \Biggr ] \; ,
    \label{eqn:dgl-weyl}
\end{eqnarray}
where $B_{i\;\mu}$ ($i=1,2,3$) and $\chi_{i}$ ($i=1,2,3$) denote the
dual gauge field and the complex scalar monopole field, respectively. 
The dual gauge fields within the Weyl symmetric expression are subject to the 
constraint $\sum_{i=1}^{3}B_{i\;\mu}=0$.  A distinctive feature of
the DGL Lagrangian is that the quark current $j_{j\;\mu}^{(e)}$ 
($j=1,2,3$) is represented as the
boundary of a nonlocal color-electric Dirac string term
$\Sigma_{j\;\mu\nu}^{(e)}$, as $j_{j \; \mu}^{(e)}=\partial^{\nu}
{}^{*\!}\Sigma_{j\;\mu\nu}^{(e)}$, which corresponds to the modified
dual Bianchi identity.  Note that $(\partial\wedge{B}_{i})_{\mu\nu}
\equiv \partial_{\mu} B_{i\;\nu}-\partial_{\nu}B_{i\;\mu}$ satisfies
$\partial^{\nu} {}^{*\!}(\partial \wedge B_{i})_{\mu\nu}=0$. }
In this framework, the color-electric charge of the quark is specified 
by using the weight vector of the SU(3) algebra, $\boldsymbol{w}_{j}$
($j=1,2,3$), as $\boldsymbol{Q}_{j}^{(e)} \equiv e \boldsymbol{w}_{j}$, 
where
$\boldsymbol{w}_1= \left (1/2, \sqrt{3}/6 \right )$, 
$\boldsymbol{w}_2= \left (-1/2, \sqrt{3}/6 \right )$, and 
$\boldsymbol{w}_3= \left (0, -1/\sqrt{3} \right )$.  On
the other hand, the color-magnetic charges of the monopole fields
$\chi_{i}$ are expressed by the root vectors of the SU(3) algebra,
$\boldsymbol{\epsilon}_i$, as $Q_{i}^{(m)} \equiv g \boldsymbol{\epsilon}_i$
($i=1,2,3$), where 
$\boldsymbol{\epsilon}_1=\left (-1/2,\sqrt{3}/2 \right )$,
$\boldsymbol{\epsilon}_2=\left(-1/2,-\sqrt{3}/2 \right )$, and
$\boldsymbol{\epsilon}_3=\left (1,0 \right )$.  
The appearance of the matrix
$m_{ij}$ in the dual field strength tensor is due to the extended
Dirac quantization condition between the color-electric and the
color-magnetic charges, $\boldsymbol{Q}_{i}^{(m)} \cdot
\boldsymbol{Q}_{j}^{(e)} = 2\pi m_{ij}$, where we have required $eg=4\pi$.
The entries of the matrix $m_{ij}$ are integers expressed by means 
of the third-rank antisymmetric tensor 
$\epsilon_{ijk}$ as $m_{ij} =2 \boldsymbol{\epsilon}_i
\cdot \boldsymbol{w}_{j} = \sum_{k=1}^{3} \epsilon_{ijk}$.  
Using the matrix $m_{ij}$, the dual gauge field is decomposed 
into two parts, $B_{i\;\mu}= B_{i\;\mu}^{\rm reg} +
\sum_{j=1}^{3}m_{ij}B_{j\;\mu}^{\rm sing}$, where the singular part
$B_{j\;\mu}^{\rm sing}$ is determined to satisfy the relation
\begin{equation}
    (\partial \wedge B_{j}^{\rm sing})_{\mu\nu}
    +2 \pi \Sigma_{j\; \mu \nu}^{(e)} 
    = 2\pi C_{j\; \mu \nu}^{(e)} \quad   (j=1,2,3) \; ,
    \label{eqn:su3-dg-singular-part}
\end{equation}
where $C_{j\; \mu \nu}^{(e)}$ is a Coulombic field which does not
contain any Dirac string,
given by~\footnote{The decomposition of the
dual gauge field is also given in Ref.~\cite{Koma:2001pz} in a more
elegant way using differential forms.}
\begin{equation}
    C_{j\; \mu\nu}^{(e)}(x) = \frac{1}{4\pi^{2}} \int d^{4}y
    \frac{1}{|x-y|^{2}}
    {}^{*\!}(\partial \wedge j_{j}^{(e)}(y))_{\mu\nu} \; .
    \label{eqn:dah-Coulomb-term}
\end{equation}
The two mass scales of the DGL theory are the mass of the dual gauge
field $m_B = \sqrt{3}gv$ and the mass of the monopole field $m_\chi =
2\sqrt{\lambda}v$.  The corresponding inverse masses are related
to the thickness of the flux tube, given by the penetration depth 
of the color-electric field into the vacuum, and to the coherence 
length of the monopole field, respectively.  
In analogy to usual superconductors, their ratio, $\kappa \equiv m_\chi/m_B$,
is labelling the type of dual superconductivity of the vacuum.

\section{Flux-tube solution}
\sloppypar{
The flux-tube solution in the cylindrical symmetric system with
translational invariance along the $z$ axis is described, as function 
of a two-dimensional radius $r$ and the azimuthal angle $\varphi$,
by the modulus of the monopole field $\phi_{i}(r) =|\chi_{i}(r)|$
and the regular part of the dual gauge 
field $B_{i}^{\rm reg}(r)~\bvec{e}_{\varphi}
= [\tilde B_{i}^{\rm reg}(r)/r]~\bvec{e}_{\varphi}$.
In this system, the contribution of the Coulomb term 
\eqref{eqn:dah-Coulomb-term} 
can set to be zero because the static charges are infinitely apart. 
Thus, Eq.~\eqref{eqn:su3-dg-singular-part} leads to
$B_{i}^{\rm sing}(r)~\bvec{e}_{\varphi}= -(n_{i}^{(m)}/r)~\bvec{e}_{\varphi}$ 
with
\begin{equation}
    n_i^{(m)} \equiv \sum_{j=1}^{3}m_{ij}n_{j}^{(e)} \; .
    \label{eqn:winding-number}
\end{equation}
Here $n_j^{(e)}$ is the winding number of $j$-type color-electric
Dirac string $\Sigma_{j\;\mu\nu}^{(e)}$, which can take various integers
depending on the representation of the SU(3) color group to which
the charges belong (see, Table~\ref{tab:values}). The
string tension of the flux tube is calculated as an energy per unit 
length in $z$ direction.
\begin{eqnarray}
    \sigma_D 
    &=& 
    2\pi     \sum_{i=1}^3   \int_0^{\infty}rdr
    \Biggl [
    \frac{1}{3g^2} \left ( \frac{1}{r}\frac{d \tilde B^{\rm reg}_{i}}
    {dr} \right )^2
    +    \left ( \frac{d \phi_i}{d r} \right)^2
     \nonumber\\*
     &&
    +
    \left ( 
    \frac{ \tilde B^{\rm reg}_{i} -  n_{i}^{(m)}}{r} 
    \right )^2 \! \phi_i^2
    +
    \lambda ( \phi_i^2 - v^2 )^2
    \Biggr ] \; .
    \label{eqn:string-tension}
\end{eqnarray}
In the Bogomol'nyi limit, $\kappa=m_\chi/m_B=1.0$, one gets the saturated
string tension analytically as~\cite{Koma:2001ut}
\begin{equation}
    \sigma_D
    =     2 \pi v^2 \sum_{i=1}^3 \left | n_{i}^{(m)} \right | 
    =    4 \pi v^{2} (p+q) \; .
    \label{eqn:st-exact}
\end{equation}
Then the ratio of the string tension between a higher and the fundamental 
representation is simply given by 
\begin{equation}
    d_{D}= \sigma_{D}/\sigma_{\bf 3}=p+q \; ,
    \label{eq:st-count}
\end{equation}
which is nothing but the sum $p+q$ of the Dynkin index of
the representation $D$ of the SU(3) group.  In the general dual
superconducting vacuum of type~I ($\kappa<1.0$) 
and of type~II ($\kappa>1.0$),
one has to evaluate the whole expression \eqref{eqn:string-tension}
in its variational minimum by solving the field equations numerically.

% % % % % % % 
\begin{table}[t]
    \setlength{\tabcolsep}{5pt}
    \centering
    \caption{Eigenvalues of the quadratic Casimir operators $C^{(2)}(D)$, 
    and of $A^{(2)}(D)$, its restriction to the Cartan subgroup, 
    for various SU(3) representations denoted by $D$ with  
    $[p,q]$ as the Dynkin index. 
    $\{n_{j}^{(e)} \}$ classifies the winding number of 
    the flux-tube solution in the DGL theory which belongs to the given 
    SU(3) representation.}
    \begin{tabular}{lcclllll}
        \hline\hline
        $D$   &  $ [p,q]$  &  $p+q$ & 
         \multicolumn{2}{l}{ $C^{(2)}(D)$ } & 
         \multicolumn{2}{l}{ $A^{(2)}(D)$ } &
         $\{n_{j=1,2,3}^{(e)} \}$\\
         & &  &  & 
         $\!\!\!\!\!$ (ratio)  &  &  
         $\!\!\!\!\!$ (ratio)  & \\
        \hline
        {\bf 3} & [1,0] & 1&  4/3 & - & 1/3 & -   & $\{1,0,0 \}$\\
        \hline
        {\bf 8} & [1,1] & 2&3 & 9/4 & 1 & 3  & $\{1,-1,0 \}$ \\
        %\hline
        {\bf 6} & [2,0] &  & 10/3 & 5/2 & 4/3 & 
        4   & $\{2,0,0 \}$\\
        \hline
        {\bf 15a} & [2,1] & 3&16/3 & 4 & 7/3 & 7   & 
        $\{2,-1,0 \}$\\
        %\hline
        {\bf 10} & [3,0] & & 6 & 9/2  & 3 & 9  & $\{3,0,0 \}$ \\
        \hline
        {\bf 27} & [2,2] &4& 8 & 6 & 4 & 12  & $\{2,-2,0 \}$ \\
        %\hline
        {\bf 24} & [3,1] & &23/3 & 23/4  & 
        13/3  & 13   & $\{3,-1,0 \}$\\
        %\hline
        {\bf 15s} & [4,0] & &28/3 & 7 & 
        16/3& 16   & $\{4,0,0 \}$\\
        \hline\hline
   \end{tabular}
    \label{tab:values}
\end{table}
% % % % % % % % 

\section{Numerical result, new features and reason}
In Fig.~\ref{fig:casimir-abelian}, we show the ratios of the string
tensions of the flux tubes, $d_{D}=\sigma_{D}/\sigma_{\bf 3}$ 
for three values of the GL parameter, $\kappa = 1.0$, $3.0$, and $9.0$
(numerically obtained for $\kappa \ne 1.0$).
We also plot  the ratios of the string tensions obtained 
by the lattice simulations of Ref.~\cite{Deldar:1999vi}.
To characterize two hypothetical cases under discussion, we plot also
the ratios of eigenvalues of the quadratic Casimir operator
evaluated in the highest weight state, 
\begin{eqnarray}
 C^{(2)}(D) &=&
\langle D_{\rm max} | (T^{3})^{2}+ (T^{8})^{2} +2 T^{3}| 
D_{\rm max} \rangle \nonumber\\
&=& 
\frac{1}{3}(p^2+pq+q^2)+(p+q) \; ,
\label{eqn:casimir}
\end{eqnarray}
as well as its Abelian-projected (Cartan restricted) values 
\begin{equation}
A^{(2)}(D) = \langle D_{\rm max} | (T^{3})^{2}+ (T^{8})^{2} 
| D_{\rm max} \rangle = \frac{1}{3}(p^2+pq+q^2) \; .
    \label{eqn:abelian}
\end{equation}
Here the relations $T^{3}| D_{\rm max} \rangle = \frac{p+q}{2} |
D_{\rm max} \rangle $ and $T^{8}| D_{\rm max} \rangle =
\frac{p-q}{2\sqrt{3}} | D_{\rm max} \rangle$ for the highest weight 
state $| D_{\rm max} \rangle$ have been used. 
Values for the lowest eight 
representations are tabulated in Table~\ref{tab:values}.
We find that the DGL result in the type-II dual superconducting 
vacuum near $\kappa = 3.0$ agrees well with all lattice data
obtained in Ref.~\cite{Deldar:1999vi}, albeit with big errors.

\par
The mechanism of the $\kappa$ dependence is understood as follows.  
In the Bogomol'nyi limit, $\kappa =1.0$, the ratio between the string 
tensions of a higher and the fundamental representation satisfies 
the {\it flux counting rule}; 
the string tension $\sigma_{D}$ is simply proportional to the number
of the color-electric Dirac strings inside the flux tube, as seen
from Eq.~\eqref{eqn:st-exact}.  
With increasing $\kappa$, the interaction ranges of these fields 
get out of balance, and an excess of energy appears because of the 
interaction between fundamental flux tubes
\cite{Jacobs:1979ch,Koma:1997qq}. This leads to systematic deviations
from the counting rule.~\footnote{In this analysis, the higher
dimensional flux tube is assumed to be stable against splitting
into fundamental ones. However there must be a certain minimal
$q$-$\bar{q}$ distance, depending on the GL parameter.
Otherwise this effect is not negligible.} 
Note that the deviation of $d_{D}$ from the counting rule grows
toward higher representations $D$, since the number of fundamental 
flux which coexist in the flux-tube of representation $D$
increases as the sum $p+q$ of Dynkin indices.

\par
On the other hand, we also find that the DGL result at $\kappa = 9.0$,
for the deeply type-II vacuum, {\it uniformly} reproduces 
{\it Casimir-like} ratios by accident,
through the deviations from the flux counting rule.
This result does not contradict the previous one~\cite{Koma:2001ut}, 
where the GL parameters were searched which reproduce the 
\emph{exact} Casimir ratio $d_D$ for each $D$.

\par
Since the DGL theory is constructed from
QCD via Abelian projection scheme, an objection has been raised 
that one would have then {\it Abelian scaling} following $A^{(2)}(D)$
~\eqref{eqn:abelian} 
instead of {\it Casimir scaling} following $C^{(2)}(D)$
~\eqref{eqn:casimir} in the DGL theory.  
Abelian scaling, for instance, would give a ratio 
between the $D={\bf 8}$ and $D={\bf 3}$ 
representations as large as 
$A^{(2)}({\bf 8})/A^{(2)}({\bf 3})=3.0$, in clear distinction from the
Casimir ratio $9/4$ {\it and} the lattice value.  
We find that the ratios of the
string tensions of flux tubes in the DGL theory 
are not steeply rising as dictated by Abelian scaling, 
although the Abelian-projected theory has been claimed to have 
Abelian scaling realized not only at short distance  
but also in the long-range
force~\cite{Shevchenko:2000du}.

% % % % % % % 
\begin{figure}[!t]
\begin{center}
    \includegraphics[ width=102mm]{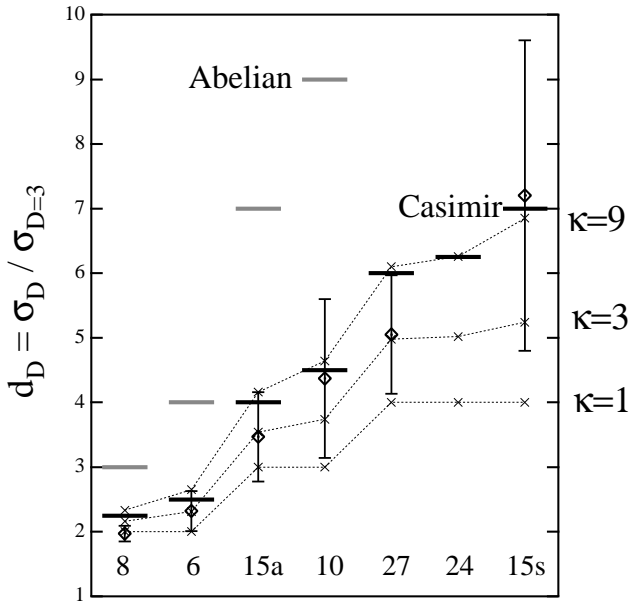}
\caption{The ratios of the string tensions of flux tubes for various
SU(3) representations, $d_{D}=\sigma_{D}/\sigma_{\bf 3}$ 
for the GL parameters $\kappa=1.0$, $3.0$ and $9.0$ (represented by 
crosses, each case connected by lines to guide the eye).  
The ratios of eigenvalues of the quadratic Casimir operators
are shown as black bars. 
Restricted to the Cartan algebra (Abelian scaling), 
the ratios are shown as gray bars.  
For comparison the lattice data of Ref.~\cite{Deldar:1999vi} are also
plotted (diamonds with error bars).}
\label{fig:casimir-abelian}
% \vspace{0.5cm}
 \end{center}
 \end{figure}
% % % % % % % % 

\section{Summary} 

We have studied the string tensions of flux tubes
associated with static charges in various SU(3) representations
in the DGL theory, based on a manifestly Weyl symmetric procedure.  
We have found that a  GL-parameter near $\kappa = 3.0$
{\it can reproduce} the ratios of string tensions
consistent with the lattice data~\cite{Deldar:1999vi}. 
We have also found that the ratios of string tensions are 
far from Abelian scaling at any finite value of $\kappa$. 
The DGL theory accidentally shows Casimir-like scaling for a deeply type-II 
vacuum with $\kappa = 9.0$. But there is no obvious relation 
to the eigenvalues of the Casimir operator.

\par
The mechanism behind the systematic behavior of string tensions 
in the DGL theory can rather be understood as a result of the flux-tube 
dynamics. This includes the possibility that the lattice data 
contain a similar dynamical effect. 
We would like to emphasize that it is important 
to have more lattice results carefully interpreted, 
without bias toward Casimir scaling, before one is able 
to judge the viability of various QCD-vacuum models.

\begin{acknowledgement}
We are grateful to E.-M. Ilgenfritz, T. Suzuki, 
H. Toki, and D. Ebert for useful discussions and comments.  
One of authors (Y.K.) acknowledges M. Faber for
interesting discussions of  Casimir scaling in the center vortex
picture.
This work is partially supported  by 
the Ministry of Education, Science, Sports and Culture,
Japan, Grant-in-Aid for Encouragement of Young 
Scientists (B), 14740161, 2002.
\end{acknowledgement}

% \bibliographystyle{h-physrev3}
% \bibliography{koma-paper}

\end{document}